\documentclass[sigconf, nonacm]{acmart}

\usepackage{amsmath,amsfonts}
\usepackage{algorithmic}
\usepackage{caption}
\usepackage{graphicx}
\usepackage{subcaption}
\usepackage{textcomp}
\usepackage{xcolor}
\usepackage{url}
\usepackage{adjustbox}
\usepackage{multirow}
\usepackage{enumitem}
\usepackage[linesnumbered,ruled]{algorithm2e}
\usepackage{bm}
\usepackage{setspace}
\usepackage{pifont}
\usepackage{xspace}
\usepackage{microtype}
\usepackage{ulem}
\usepackage{kotex}

\newcommand\vldbdoi{XX.XX/XXX.XX}

\newcommand\vldbavailabilityurl{URL_TO_YOUR_ARTIFACTS}
\newcommand\vldbpagestyle{plain} 

\newcommand\noh[1]{ {\color{purple}{#1}} }
\newcommand{\cnoh}[1]{{\color{violet}{\textit{#1}}}} 
\newcommand{\dnoh}[1]{{\color{violet}{\sout{#1}}}} 

\newcommand\inho[1]{ {\color{olive}{#1}} }
\newcommand{\cinho}[1]{{\color{teal}{\textit{#1}}}}

\newcommand\hb[1]{ {\color{blue}{#1}} }

\begin{document}
\title{A CXL Memory Rack for Multi-Turn LLM Serving}

\newcommand{\CXLHM}{CXL-HM\xspace}
\newcommand{\sysname}{HyMCache\xspace}
\newcommand{\sqnum}[1]{%
  \colorbox{blue}{\textcolor{white}{\scriptsize\bfseries #1}}%
}

\definecolor{VIOLET}{rgb}{0.56,0.0,1.0}

\author{%
Hakbeom Jang$^{1}$,
Inho Song$^{1,2,\dagger}$,
Hoshik Kim$^{1}$,
Sam H. Noh$^{2}$,
Jongryool Kim$^{1}$%
}

\affiliation{%
  \institution{%
    $^{1}$SK hynix America \quad
    $^{2}$Virginia Tech
  }
}

\renewcommand{\shortauthors}{Jang et al.}


\begin{abstract}

Long-context, multi-turn, and agentic LLM workloads increasingly reuse previously processed context, making KV-cache reuse essential for reducing redundant computation. However, this reuse shifts the bottleneck to the memory tier that stores and serves reusable KV states at cluster scale. GPU HBM and host DRAM are too costly to scale to TB-scale shared context capacity, motivating remote tiers built from lower-cost, higher-capacity media.
This paper presents \sysname, a CXL memory rack for multi-turn LLM serving. We build the memory rack using cost-efficient CXL-hybrid memory (\CXLHM), which combines a small amount of in-device DRAM with large SSD-backed capacity behind a CXL interface. By exploiting the read-dominant, predictable, and append-only nature of multi-turn KV-cache access, \sysname rethinks DRAM management within \CXLHM to efficiently support TB-scale SSD-backed KV reuse. It uses request-level prefix prefetching and opportunistic write buffering to stage latency-critical reads in device DRAM, enabling DRAM-scale KV-cache efficiency at SSD-level cost.
We evaluate \sysname on a real \CXLHM prototype under both single-aggregator and PD-disaggregated serving configurations. 
Under the same DRAM budget, \sysname{} outperforms local LMCache by 3.0$\times$ in single-node serving and 1.45$\times$ in PD-disaggregated serving. Compared with 1~TB distributed-DRAM Mooncake, \sysname{} incurs about 30\% lower performance, but uses 16$\times$ less DRAM. 

\end{abstract}

\maketitle
\begingroup
\renewcommand{\thefootnote}{\fnsymbol{footnote}}
\footnotetext[2]{This work was done during the second author's internship at SK Hynix.}
\endgroup

\pagestyle{\vldbpagestyle}


\section{Introduction}

LLM serving is becoming increasingly memory-dominated. As workloads evolve from single-turn prompts to long-context, multi-turn, and agentic applications~\cite{openai_chatgpt_agent,anthropic_claude_computer_use,google_gemini_long_context,deepseek_context_cache}, inference cost is shaped not only by computation, but also by the memory capacity~\cite{kwon2023vllm,cheng2025lmcache,qin2025mooncake,sheng2023flexgen,infinigen-osdi-2024}
to retain and reuse growing context state.
The memory-capacity pressure created by modern LLM serving 
translates into expensive GPU and DRAM provisioning
and lower serving efficiency. 
Specifically, KV cache
reuse reduces repeated context-processing work by retaining the intermediate state generated for previously processed tokens, but it also turns the KV cache into reusable data whose capacity, movement, and sharing increasingly influence serving efficiency across turns, sessions, and workers.

This paper presents \sysname, a CXL memory rack for multi-turn LLM serving.
We build \sysname using cost-efficient CXL-hybrid memory, hereafter also
referred to as \CXLHM, which combines a small amount of in-device DRAM
with large SSD-backed capacity. By integrating memory-addressable
\CXLHM devices into a shared serving tier, \sysname enables DRAM-scale
KV-cache efficiency at SSD-level cost.

The memory-capacity pressure is already pushing LLM serving toward shared, storage-based context infrastructure. TB-scale and beyond reusable context capacity is difficult to support economically with HBM or DRAM alone, so publicly disclosed frontier systems are beginning to incorporate lower-cost, 
high-capacity storage media into the inference memory hierarchy. NVIDIA CMX (Context Memory Storage) introduces a dedicated context-memory tier for long-context and multi-agent inference~\cite{nvidia_icms_2026}, while DeepSeek Context Caching on Disk shows that disk-backed context reuse is being adopted in commercial LLM APIs~\cite{deepseek_context_cache}. These efforts show that KV-cache reuse is increasingly becoming a memory-tiering problem.
Storage media provide the capacity and cost efficiency needed to retain large reusable contexts, while optimizations such as fast staging, prefetching, and data-movement control are required to meet the latency demands of context reuse in LLM serving systems.


To build such a shared context layer, the ecosystem is exploring both storage-centric and memory-centric designs. On the storage side, networking and DPU vendors such as NVIDIA~\cite{nvidia_bluefield_dpu,nvidia_doca_nvmeof}, Broadcom~\cite{broadcom_stingray_smartnic}, Marvell~\cite{marvell_octeon_dpu,marvell_nvmeof}, Xsight Labs~\cite{xsight_e1_dpu}, and Chelsio~\cite{chelsio_unified_wire} are combining RDMA/NVMe-oF, SmartNICs, and DPUs with dense SSDs from storage vendors such as SK hynix~\cite{skhynix_ps1000_essd} and Samsung~\cite{samsung_pm1743} to build remote flash tiers~\cite{sun2025scalio, guo2023leed, gimbal_sigcomm21, zhan2024tricklekv}. On the memory side, memory vendors, cloud providers, and startups, including SK hynix~\cite{skhynix_cmm_ddr5,skhynix_niagara2},
Samsung~\cite{samsung_cmm_d,samsung_cmm_b},  
Alibaba~\cite{alibaba_cxl_distributed_memory}, Liqid~\cite{liqid_composable_memory}, UniFabriX~\cite{unifabrix_smart_memory_node}, and XCENA~\cite{xcena_mx1} are developing CXL-based memory expansion, rack-scale memory pooling, and CXL memory appliances. 

Despite this broad momentum, the most practical architecture remains unclear. Storage-backed tiers provide large capacity at lower cost, but still require storage-node software to stage SSD-resident KV blocks through DRAM buffers before remote access. DPU-based offload can reduce part of this overhead, but adds software complexity and non-trivial appliance cost. In contrast, DRAM-based memory pools provide a cleaner low-latency memory abstraction, but scaling them to tens or hundreds of terabytes is costly.

As a middle ground,
major memory vendors such as SK hynix~\cite{jang2026itmeinferencetieredmemory} and Samsung~\cite{samsung_cmmh_hotstorage} have proposed CXL-hybrid memory, or \CXLHM, as a compromise between DRAM-only memory expansion and SSD-based capacity expansion. 
\CXLHM exposes the SSD-backed capacity to the host through a memory interface (i.e., CXL.mem), while using the device-side DRAM transparently as an internal cache.
This makes it an attractive substrate for remote KV caching: the SSD-backed region provides TB-scale capacity at lower cost, while the internal DRAM can serve latency-critical accesses when managed appropriately.

However, we find that generic internal resource management including cache-management policies, such as the LRU policy used in commercial \CXLHM devices, are insufficient for agentic LLM serving. We therefore design and prototype a \CXLHM tailored to the workload characteristics of multi-turn, agentic LLM serving. To exploit the read-dominant, predictable, and cross-turn reuse patterns of these workloads, \sysname provides an API that allows the serving stack to prefetch KV blocks into \CXLHM’s internal DRAM, thereby hiding retrieval latency from the SSD-backed region.
We evaluate \sysname under both single-aggregator and Prefill-Decode (PD) disaggregated serving configurations. With the same DRAM budget, \sysname{} improves serving performance over local LMCache by 3.0$\times$ in single-node serving and 1.45$\times$ in PD-disaggregated serving with 64~GB per worker, 256~GB in total. Compared with 1~TB distributed-DRAM Mooncake, \sysname{} trades a 30\% performance gap for a 16$\times$ reduction in DRAM usage.

 \section{Background and Motivation}

\subsection{LLM Serving and KV-Cache Tiers}

\noindent\textbf{LLM inference and KV cache.}
LLMs generate output tokens autoregressively: each new token is produced based on the input prompt and all previously generated tokens. In modern LLM serving systems, inference is commonly divided into two phases, prefill and decode~\cite{kwon2023vllm}. During prefill, the model processes the input prompt and materializes key/value (KV) tensors for all prompt tokens. During decode, the model generates one token at a time and attends to the KV tensors produced for earlier tokens. The KV cache stores these intermediate tensors so that later attention computation can reuse them instead of regenerating them. Within a request, this avoids recomputing KV tensors during decode; across requests or turns, it also enables reusable prefixes to skip repeated context processing.

\noindent\textbf{LLM serving stack.}
Modern LLM serving stacks combine a GPU-side serving engine with a distributed serving framework. Serving engines such as vLLM~\cite{kwon2023vllm} manage request batching, scheduling, and GPU KV-cache allocation for high-throughput inference. vLLM organizes the GPU KV cache into fixed-size blocks and uses PagedAttention to improve memory utilization. Distributed frameworks such as Dynamo~\cite{dynamo} build on these runtimes to support multi-worker serving, including Prefill--Decode (PD) disaggregation, request routing, and inter-worker KV transfer. In PD-disaggregated serving, prefill workers process input contexts and produce KV blocks, while decode workers continue token generation after receiving the required KV blocks. Dynamo uses NIXL~\cite{nixl} for the P--D KV-transfer path, and NIXL can use transport backends such as UCX~\cite{shamis2015ucx} to run over RDMA-capable networks in cross-node deployments.

\noindent\textbf{KV-cache tiers.}
KV-cache reuse is commonly implemented as a tiered memory hierarchy. One useful framing is NVIDIA's G1--G4 context hierarchy~\cite{nvidia_icms_2026}: G1 is GPU HBM, where serving engines such as vLLM manage resident KV blocks through GPU KV-cache allocation and prefix caching; G2 is system DRAM, used by systems such as LMCache and Mooncake to extend KV capacity within or across workers~\cite{cheng2025lmcache,qin2025mooncake}; G3 is local SSD, used as a lower-cost capacity tier through KV offloading or DRAM--SSD caching; and G4 is shared remote storage, where larger context data can be accessed through storage/network fabrics such as NVMe-oF, RDMA, or DPU-assisted data paths. Recent designs also introduce intermediate tiers and optimizations between these levels, such as G2.5 or G3.5 context-memory tiers, to bridge the latency--capacity gap between DRAM and storage. Overall, KV-cache tiering exposes a trade-off among latency, capacity, cost, energy, and data movement.


\subsection{CXL Solutions for Building a Memory Rack}

Compute Express Link (CXL) has emerged as a memory interconnect that extends server memory capacity beyond local DIMMs while preserving a memory-like access model. The CXL specification has evolved from single-device attachment in CXL~1.1 to switch-based memory pooling in CXL~2.0 and fabric-oriented multi-level switching in CXL~3.0. Its sub-protocols separate device management and I/O semantics through CXL.io, cache-coherent accelerator access through CXL.cache, and load/store memory access through CXL.mem~\cite{wang-cxl-realworld-arxiv24}.

CXL creates a distinct latency--capacity point for KV-cache tiering, sitting much closer to DRAM than to storage while allowing memory capacity to scale beyond local DRAM. Importantly, commercial CXL platforms provide memory access latencies comparable to conventional NUMA systems. Local DRAM access is typically around 80--140~ns~\cite{almaruf2023tpp}, remote NUMA access is roughly 2$\times$ higher, and CXL memory expansion modules have been reported to be around 170--250~ns~\cite{almaruf2023tpp,sth_cxl_latency,nextplatform_cxl_latency}. More recently, commercial CXL~2.0 switches have expanded the practical scope of CXL beyond single-node memory expansion toward large-scale shared memory pools. For example, the XConn (acquired by Marvell) CXL switch supports up to 256 lanes~\cite{xconn_cxl_switch}, while prior work reports roughly 750~ns minimal 64-byte I/O latency for an XConn-based CXL switch fabric~\cite{beluga_cxl_kvcache}. Pooled-memory prototypes such as SK hynix Niagara further show that shared CXL memory can remain sub-microsecond, reporting 600~ns access latency~\cite{skhynix_niagara2}.

\begin{figure}[t]
\centering
\includegraphics[width=0.48\textwidth]{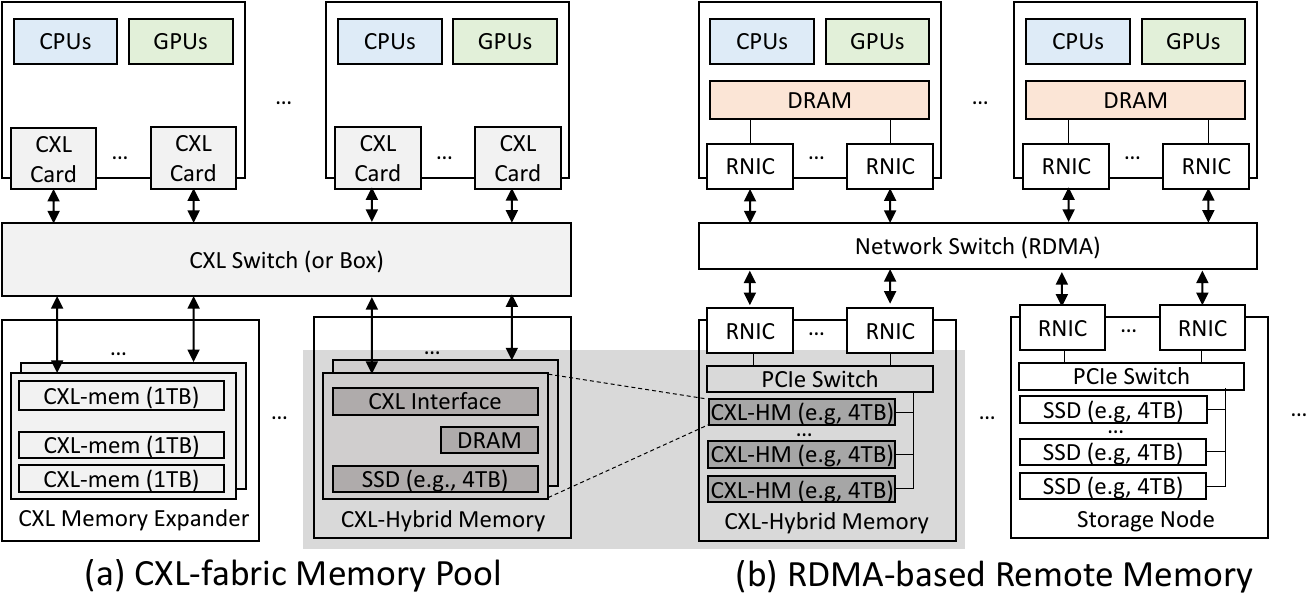}
\caption{CXL memory rack deployment models: CXL-fabric access and
RDMA-based remote access.}
\label{fig:cmm-infra}
\vspace{-3mm}
\end{figure}

The design of CXL memory modules can vary significantly depending on
system and application requirements. To build a CXL memory rack for
large shared KV-cache tiers, existing CXL memory solutions can be
broadly organized into three design points: DRAM-based CXL memory
expansion, switch-based CXL memory pooling, and \CXLHM.

A CXL memory expander is a DRAM-based CXL memory module that targets latency- and bandwidth-sensitive workloads by expanding memory capacity beyond local DIMM channels while preserving memory-semantic access~\cite{skhynix_cmm_ddr5,samsung_cmm_d}. For KV-cache reuse, such devices can provide a low-latency capacity tier for KV blocks that no longer fit in GPU memory but still require fast access. A CXL switch-based memory pool extends this model to rack-scale environments by aggregating multiple CXL memory devices into a pooled memory appliance, allowing memory capacity to be flexibly allocated and shared across hosts~\cite{samsung_cmm_b,skhynix_niagara2,beluga_cxl_kvcache}. This model is attractive for distributed LLM serving because reusable KV states can be shared across workers rather than being replicated in each server. In contrast, \CXLHM combines internal DRAM with high-capacity SSD, exposing SSD-backed capacity through a memory interface while using internal DRAM as a staging/cache layer~\cite{samsung_cmmh_hotstorage,jang2026itmeinferencetieredmemory}. This design point is especially relevant for building a large-capacity
memory rack for LLM serving because it provides much larger capacity
than DRAM-only CXL devices at lower cost.

\begin{table}[t]
\centering
\caption{Cost comparison for 15.36~TB memory and storage capacity~\cite{kioxia_cd8p_price_2026, micron7450_price_2026, samsung_pm1743_price_2026, serversupply_128gb_rdimm_2026, serversupply-solidigm-p5430, nemix_256gb_rdimm_2026, terasic_agilex7_iseries_price_2026}.}
\label{tab:cost_comparison}
\small
\setlength{\tabcolsep}{6pt}
\begin{tabular}{llcc}
\toprule
\textbf{Media} & \textbf{Config.} & \textbf{Cost} & \textbf{Cost Ratio} \\
\midrule
DRAM & 120$\times$128GB DDR5 & $\sim$\$782K & 57.5$\times$ \\
DRAM & 60$\times$256GB DDR5  & $\sim$\$815K & 59.9$\times$ \\
\midrule
NAND & Gen5 TLC (Base)       & $\sim$\$13.6K & \textbf{1.0$\times$} \\
NAND & Gen5 QLC              & $\sim$\$3.0K  & 0.22$\times$ \\
NAND & Gen4 TLC              & $\sim$\$4.8K  & 0.35$\times$ \\
NAND & Gen4 QLC              & $\sim$\$3.0K  & 0.22$\times$ \\
\midrule
\textbf{CXL-HM}$^{\dagger}$ 
     & Gen5 TLC
     & $\sim$\$40K
     & $\sim$2.9$\times$ \\
\bottomrule
\end{tabular}

\vspace{0.3em}
\small
$^{\dagger}$The \CXLHM{} cost is a rough estimate for four FPGA-based prototypes.
\end{table}

Recent literature and industry efforts have explored remote KV-cache tiers based on DRAM-backed CXL memory expansion and pooling, as illustrated in Figure~\ref{fig:cmm-infra}(a). Such systems aggregate multiple CXL memory expanders through a CXL switch, memory box, or fabric, allowing hosts to dynamically expand and share memory capacity. This direction is increasingly relevant for LLM serving, where large KV-cache states are retained and reused across requests to improve throughput and reduce redundant computation~\cite{nvidia_llm_inference_optimization,sglang_radixattention,kwon2023vllm,cheng2025lmcache,qin2025mooncake}. It has been pursued by memory vendors, cloud providers, startups, and system vendors, including SK hynix~\cite{skhynix_cmm_ddr5,skhynix_niagara2}, Samsung~\cite{samsung_cmm_d,samsung_cmm_b}, Alibaba~\cite{alibaba_cxl_distributed_memory}, Liqid~\cite{liqid_composable_memory}, UniFabriX~\cite{unifabrix_smart_memory_node}, and XCENA~\cite{xcena_mx1}, as well as research systems such as TraCT and Beluga~\cite{tract_cxl_kv_cache,beluga_cxl_kvcache}.
However, building a large CXL memory rack entirely from DRAM-based
CXL memory expanders remains costly at scale. Although these designs provide high-performance memory-semantic access, Table~\ref{tab:cost_comparison} shows that provisioning a DRAM-only tier at roughly 15~TB capacity requires hundreds of thousands of dollars in RDIMM cost.

An alternative deployment path is to place CXL-HM behind an existing RDMA-capable datacenter network, as shown in Figure~\ref{fig:cmm-infra}(b). 
In this model, the CXL-HM device is attached to a remote node, while LLM workers access reusable KV blocks through RDMA rather than through a CXL fabric. This deployment does not require CXL switches or CXL fabric support on every worker node, making it more compatible with existing cluster infrastructure. Unlike conventional remote storage, the CXL-HM node exposes SSD-backed capacity as a memory-addressable tier, allowing workers to access KV blocks through RDMA 
operations rather than storage I/O.

For \CXLHM, 
even with the FPGA/controller overhead, 
Table~\ref{tab:cost_comparison} 
illustrates the potential of \CXLHM to expose large SSD-backed capacity through a memory interface without provisioning the same amount of DRAM. As reusable KV states scale to tens or hundreds of terabytes across serving clusters, the cost
overhead of DRAM-only tiers becomes difficult to ignore. This motivates using \CXLHM to build a large-capacity CXL memory rack
for LLM serving, retaining a memory interface while using SSD-backed
capacity to reduce the cost of rack-scale memory.

\subsection{\CXLHM for Multi-Turn KV-Cache Tiering}
\CXLHM provides an opportunity to build large-capacity remote memory tiers at lower cost than DRAM-only CXL memory pools. Unlike DRAM-based CXL memory expanders, which are composed entirely of DRAM, CXL-HM combines a small amount of internal DRAM with high-capacity SSDs. CXL-HM exposes SSD-backed capacity as a host-addressable memory tier through a memory-semantic interface (i.e., CXL.mem), while internally staging data between internal DRAM and the SSD-backed region. As a result, CXL-HM can expose large SSD-backed capacity as a CPU-less NUMA node and allow applications to use it as expanded memory.

\begin{figure}[t]
\centering
\includegraphics[width=0.48\textwidth]{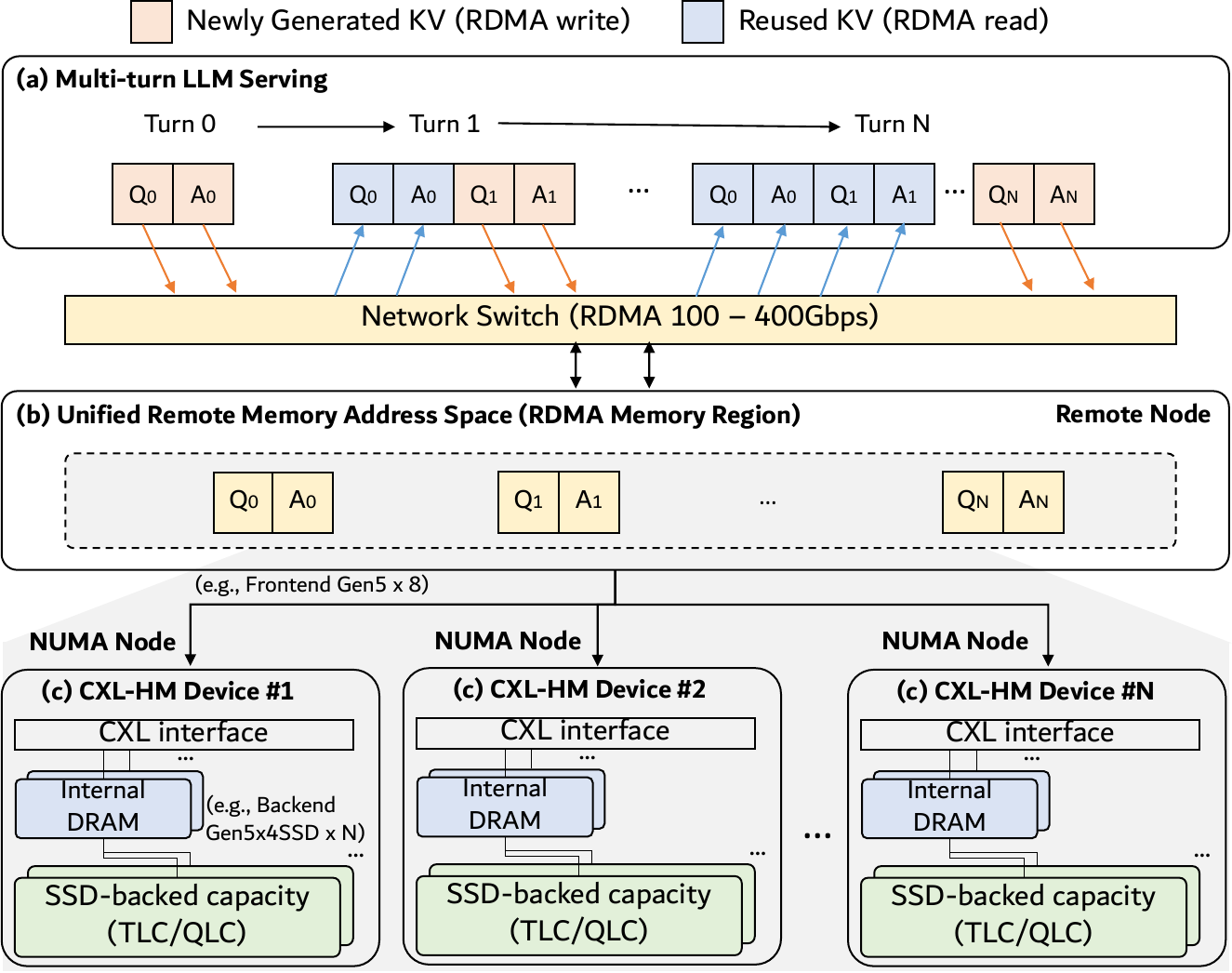}
\caption{CXL-HM opportunity for multi-turn LLM serving.} 
\label{fig:cmm-h_oppt}
\end{figure}

Figure~\ref{fig:cmm-h_oppt} illustrates an example deployment of CXL-HM as a remote memory tier for multi-turn LLM serving. The key observation is that this memory abstraction naturally matches the access pattern of multi-turn workloads: as shown in Figure~\ref{fig:cmm-h_oppt}(a), prefix KV blocks generated in earlier turns are repeatedly reused by later requests, while newly generated KV blocks are appended after each turn. In a remote serving configuration, workers can fetch reused prefix KV blocks through RDMA reads and store newly generated KV blocks through RDMA writes, allowing KV reuse to be handled through a remote memory rather than a storage-oriented path.

To scale capacity beyond a single device, the CXL-HM node manages multiple CXL-HM devices as a unified remote KV address space, as illustrated in Figure~\ref{fig:cmm-h_oppt}(b). Each device is exposed as a CPU-less NUMA node backed by its SSD capacity, and the CXL-HM node aggregates these NUMA regions into one remote memory tier visible to LLM workers. Figure~\ref{fig:cmm-h_oppt}(c) shows the internal access path of each CXL-HM device: the SSD-backed region provides the large visible capacity, while internal DRAM is used transparently as an internal cache. DRAM hits are served through the fast path, whereas misses require staging data from SSD into DRAM before completion. Therefore, the effectiveness of CXL-HM depends critically on managing this limited internal DRAM to keep predictable, latency-critical KV reuse on the fast path.

In summary, the opportunities \CXLHM brings can be summarized as follows:

\begin{itemize}
\item \textbf{Cost-efficient SSD-backed capacity.}
By relying on high-density TLC or QLC NAND SSDs attached through commodity PCIe Gen4/Gen5 NVMe interfaces, CXL-HM can provide TB-scale capacity at much lower cost per GB than DRAM-only CXL memory pools.

\item \textbf{Exploiting predictable KV reuse in CXL-HM.}
Recent agentic LLM traces show that runs can span dozens to hundreds of turns, grow up to one million tokens, and reuse more than 95\% of tokens across rounds~\cite{dualpath_agentic_kv}. This creates a highly predictable, read-heavy prefix KV access pattern: most reused KV blocks are known before execution, while only newly appended tokens generate new KV states. This predictability enables coordinated prefetching and internal DRAM staging in CXL-HM.
\end{itemize}


\begin{figure}[t]
\centering
\includegraphics[width=0.48\textwidth]{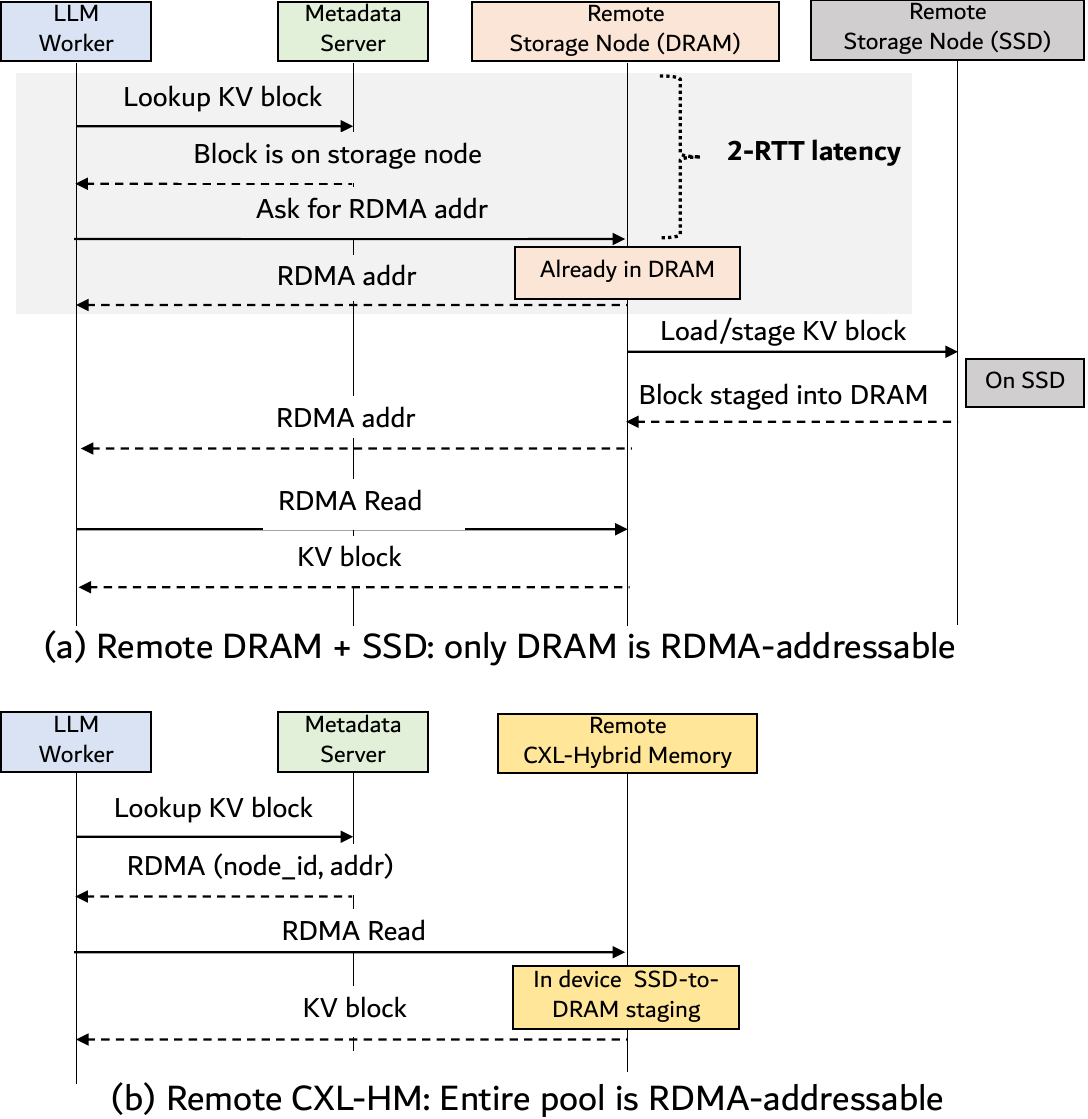}
\caption{Comparison of RDMA-based remote access paths.
}
\label{fig:cxl-hybrid-comm-overhead}
\end{figure}

\subsection{CXL-HM vs. DRAM--SSD tiering}

At this point, it is useful to clarify the difference between using \CXLHM and building a remote tier from conventional DRAM and SSDs, since the two approaches rely on similar underlying components.
Figure~\ref{fig:cxl-hybrid-comm-overhead} compares 
the remote access paths of the two approaches.  
To read a remote KV block, for both approaches, an LLM worker first queries the metadata server to identify the storage node containing the requested block. 
Then, in conventional DRAM--SSD tiering, the request is forwarded to the storage node, which performs a second metadata lookup to determine whether the block resides in DRAM or SSD. If the block is cached in DRAM, the corresponding RDMA-accessible address is returned. Otherwise, the block is first staged from SSD into an RDMA-accessible DRAM buffer, which will elongate the response, before the newly allocated DRAM address is returned to the worker. In contrast, CXL-HM exposes the entire SSD-backed capacity as a host-visible memory space. Therefore, the metadata server directly returns the final $\langle node_id, addr\rangle$ for the requested KV block, allowing the worker to immediately issue RDMA reads or writes without additional storage-node address resolution or software-mediated staging. This shortens the control path and simplifies metadata management.





\section{
\CXLHM for LLM serving 
}

This section presents our design of \CXLHM, a key component we develop for \sysname. Unlike general-purpose CMM-H products, our \CXLHM is designed specifically for multi-turn LLM inference. We therefore begin in Section~\ref{sec:characterizing-multiturn-workload} by characterizing the relevant workload properties found in multi-turn LLM inference. Based on this analysis, Section~\ref{sec:cxlhm-design-directives} derives the design directives for \CXLHM, followed by a discussion of the \CXLHM prototype developed for this study in Section~\ref{sec:LLM-targetted-CXLHM}.
\subsection{Characterizing Multi-Turn Workloads}
\label{sec:characterizing-multiturn-workload}

Multi-turn and agentic LLM serving workloads frequently reuse expanding prompt prefixes~\cite{kwon2023vllm,sglang_radixattention,cheng2025lmcache,qin2025mooncake,dualpath_agentic_kv}. Prior work reports that agent trajectories often span dozens to hundreds of turns, allowing more than 95\% of the tokens to be reused across turns~\cite{dualpath_agentic_kv}. When these prefix states are distributed across a hierarchical KV cache, the execution bottleneck shifts from compute-heavy prefill operations to read-intensive metadata lookups and block-loading loops spanning GPU memory, host DRAM, and remote tiers.

To understand these workload characteristics, we analyze KV-cache accesses in an idealized remote-DRAM setting. Our setup consists of a single aggregator with one NVIDIA A100 GPU worker connected to a 256~GB remote-DRAM tier over a 100~Gbps RDMA path. To expose remote KV-cache behavior, we intentionally constrain the GPU-side KV-cache capacity, forcing accumulated prefix KV blocks to be served from remote DRAM rather than remaining resident in GPU memory.

For the workload, we use Llama-3.1-8B with vLLM's 128-token KV block size. Even this relatively small model generates 16~MB KV blocks, while larger models produce substantially larger blocks, such as 32~MB for Qwen2.5-32B and 168~MB for OPT-30B. As a result, KV-cache movement manifests as large, block-granular transfers rather than small cache-line accesses. We run chat-based multi-turn requests from the LMSYS dataset~\cite{zheng2024lmsyschat1m} to study how the remote KV-cache footprint grows across turns.

\begin{figure}[tb]
\centering
\includegraphics[width=0.45\textwidth]{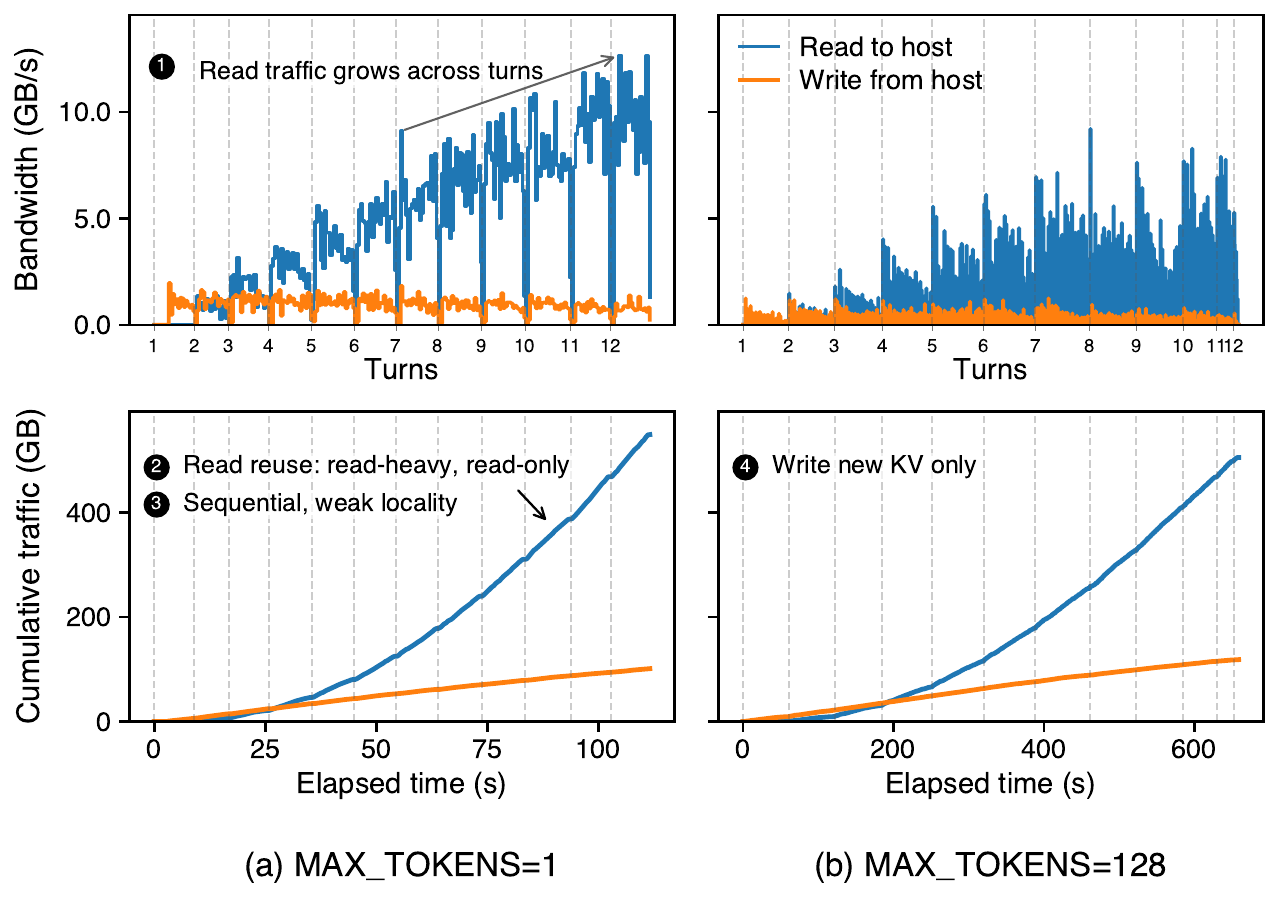}
\caption{{RDMA bandwidth from one prefill node to the remote-DRAM-only tier }
}
\label{fig:remote-dram-bw}
\end{figure}

Figure~\ref{fig:remote-dram-bw} shows aggregate KV-cache traffic generated by a single prefill node running 512 concurrent requests across multiple conversation turns, under two decode lengths: \texttt{MAX\_TOKENS}=1 and \texttt{MAX\_TOKENS}=128.
This traffic reflects a single prefill node. In deployments where multiple prefill nodes share the same remote tier, the aggregate demand on that tier can increase roughly in proportion to the number of attached prefill nodes. We highlight four read- and write-side characteristics of this traffic.

\begin{itemize}
\item \textbf{Observation \ding{202}: Read traffic grows across turns.}
As the accumulated context grows, each turn needs to reload more prefix KV blocks. 
This causes both instantaneous read traffic and cumulative read volume to increase over turns.

\item \textbf{Observation \ding {203}: KV reuse is read-heavy and read-only.}
Previously generated KV blocks are reused as prefix state and read back in later turns, but they are not modified during reuse. From the perspective of the memory tier, these KV objects remain clean, that is, they are not updated, and only serve read requests.


\item \textbf{Observation \ding{204}: Sequential KV reads with weak locality.}
While not directly observable in Figure~\ref{fig:remote-dram-bw}, we find that KV blocks are read sequentially in context order, and this sequential access yields little short-term reuse, that is, most blocks are read once within a turn and are rarely revisited soon after.
This pattern resembles a one-hit-wonder workload~\cite{yang2023s3fifo}, a scan-like access stream that pollutes the cache and neutralizes recency-based locality.
For example, in our setup the prefix-read footprint added between Turn~6 and Turn~7 alone exceeds 64~GB. 
As each turn extends the context, the KV-cache footprint grows monotonically with conversation length. As a result, increasing the DRAM-cache capacity only defers, rather than eliminates, cache pollution.
Such scan-like access pattern inherently lacks the temporal reuse that recency-based caching policies rely on.
This effect is amplified when multiple prefill nodes share the same remote tier, since their long prefix reads interleave and further dilute whatever locality remains.
\item \textbf{Observation \ding{205}: Writes are append-only of newly generated KV blocks.} Each turn writes only the KV blocks produced for its newly generated output tokens. As a result, when output lengths are similar, write volume remains relatively stable across turns. 

\end{itemize}

\begin{figure}[t]
\centering
\includegraphics[width=0.48\textwidth]{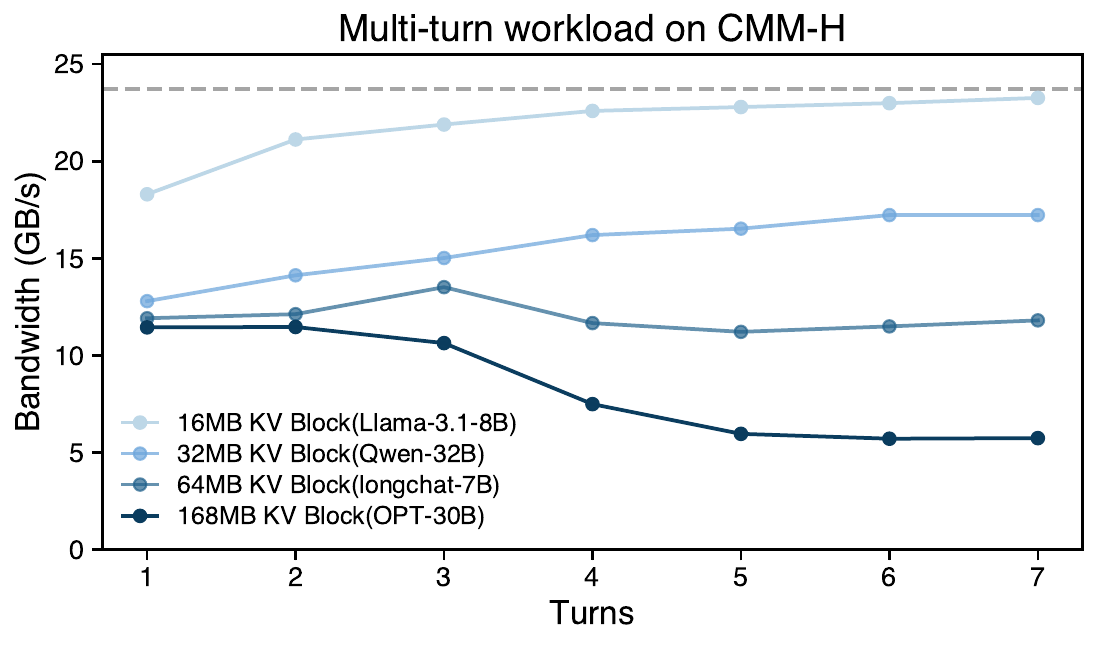}
\caption{Impact of model-dependent KV block size on CMM-H bandwidth. Results cover representative KV block sizes from different LLMs, ranging from 16~MB to 168~MB. Dashed gray line for RDMA 200Gbps.}
\label{fig:lru-lfu-analysis}
\end{figure}

\subsection{\CXLHM under Multi-Turn KV Traffic}
\label{sec:cxlhm-design-directives}


CXL-HM is an attractive substrate for a cost-efficient remote memory tier, but its performance depends critically on how device-side DRAM is managed. Existing CXL-HM prototypes and 
products, typically referred to as CMM-H devices, normally
use this DRAM as a transparent cache for SSD-backed capacity, governed by LRU-like replacement policies~\cite{samsung_cmmh_hotstorage, xcena_mx1, jang2026itmeinferencetieredmemory}.
This choice is reasonable for generic memory expansion, but it is a poor match for the long-context and agentic LLM serving access pattern 
observed in Section~\ref{sec:characterizing-multiturn-workload}.
As was shown,
these workloads repeatedly scan the accumulated prefix across turns, generating large, predictable reads with little short-term locality. Consequently, under LRU, the limited DRAM is populated with already-consumed KV blocks, while the prefix blocks needed in the next turn remain in the SSD-backed region.

To validate this behavior, we run a multi-turn KV-cache microbenchmark on a commercial CMM-H device whose 256~GB of device-side DRAM is 
managed as a transparent LRU cache for various representative KV block sizes. In these experiments, a remote worker accesses the CMM-H address space over a 200~Gbps RDMA path, issuing 64 concurrent requests that each append 32 KV blocks per turn. Turn~1 writes only the newly generated blocks. From Turn~2 onward, each request first reads the prefix written in previous turns and then appends its new blocks. Thus, the per-turn write footprint scales with KV-block size, from 32~GB for 16~MB blocks to 336~GB for 168~MB blocks, while read footprints grow monotonically across turns. Figure~\ref{fig:lru-lfu-analysis} shows the results, which we now discuss.



We observe that as long as the active footprint fits in device-side DRAM, bandwidth remains close to the RDMA line rate, as shown for Llama-3.1-8B.
However, as the footprint grows beyond the effective DRAM-cache capacity, bandwidth begins to collapse toward the SSD-backed regime, reaching roughly 5~GB/s under mixed reads and writes.
In Figure~\ref{fig:lru-lfu-analysis}, this collapse occurs between Turns~2 and~3 for 168~MB blocks and between Turns~3 and~4 for 64~MB blocks. For smaller blocks, the same effect is delayed, but once the accumulated footprint eventually fills the cache, bandwidth is expected to degrade in the same manner.

We attribute this collapse to the conventional cache-management policies used in CMM-H, particularly LRU-induced refills and dirty evictions, which we verify internally and which are consistent with prior CMM-H microbenchmark results~\cite{samsung_cmmh_hotstorage}. LRU-induced refills cause one-shot-wonder blocks to remain in the cache even after they have been consumed, occupying space that could otherwise hold upcoming prefix blocks that we know, from Section~\ref{sec:characterizing-multiturn-workload}, will soon be needed. Dirty eviction is more subtle. As discussed above, write traffic is stable and bounded, but newly written KV blocks are quickly propagated to persistent flash media according to the device’s internal policy to protect against data loss. Consequently, read traffic becomes interleaved with backend writebacks inside the CMM-H device, degrading the effective read bandwidth.

Thus, our central challenge is to design \CXLHM for \sysname to overcome the limitations of conventional CMM-H devices and directly support the observed workload characteristics of multi-turn LLM serving.

\subsection{LLM Targeted \CXLHM}
\label{sec:LLM-targetted-CXLHM}

The workload observations above motivate a different way of using \CXLHM for LLM serving. Rather than treating the internal DRAM as a transparent LRU cache, our prototype uses it as an explicitly managed staging space for KV objects. The key idea is to use the predictable KV-object access order revealed by multi-turn prefix-cache lookup, allowing \CXLHM to stage upcoming KV objects before the foreground remote reads arrive.

\begin{figure}[t]
\centering
\includegraphics[width=0.48\textwidth]{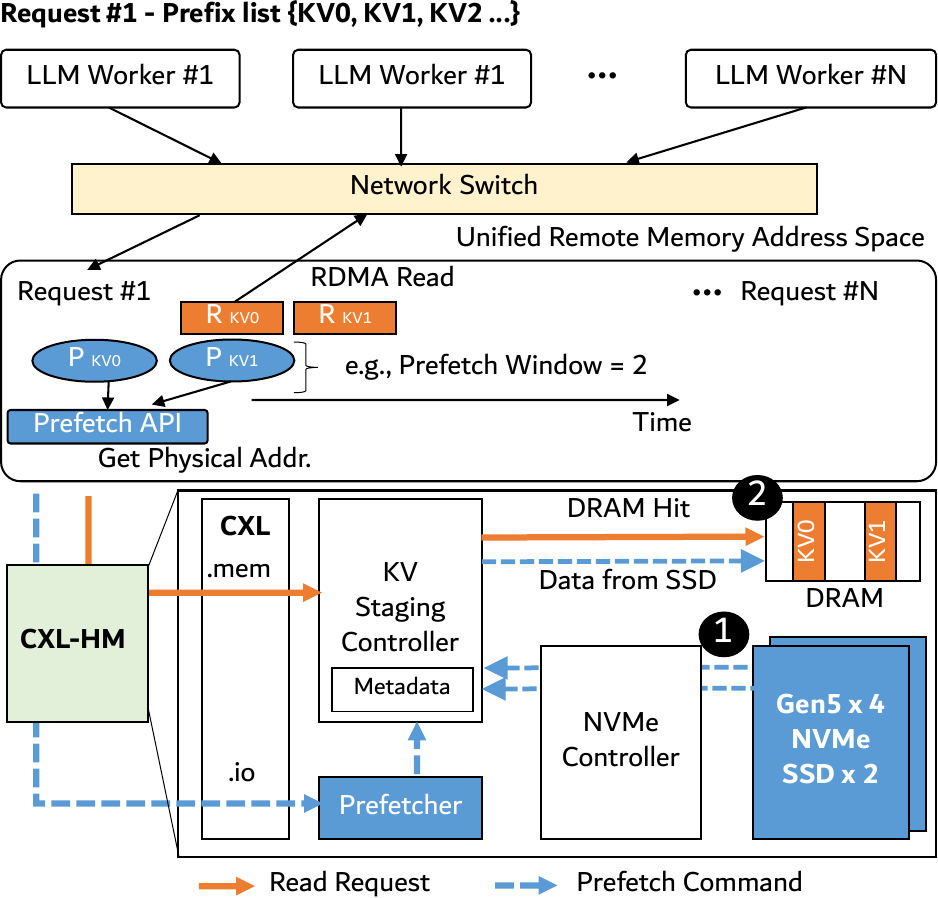}
\caption{Overview of \CXLHM prototype and its interaction with the workload.
}
\label{fig:proposed_arch}
\end{figure}

Figure~\ref{fig:proposed_arch} shows the overview of our LLM-targeted \CXLHM prototype. LLM workers access the remote KV tier as remote memory through RDMA or a CXL switch-based fabric; our prototype uses RDMA as the example path. The remote tier is exposed as a unified memory address space backed by one or more \CXLHM devices. 
A read request is served directly from internal DRAM if the requested KV object is already staged; otherwise, the object is first copied from SSD-backed capacity into internal DRAM before being returned. The prototype focuses on two design points for LLM serving: KV-object prefetching and latency-hiding staging window.
\\

\noindent\textbf{\ding{202} KV-object prefetching.}
Multi-turn prefix-cache lookup reveals the ordered list of KV objects that an inference request will consume. The serving runtime passes this object-level access order to \CXLHM, allowing upcoming KV objects to be prefetched before foreground remote reads arrive. \CXLHM provides an explicit prefetch interface for user-level control of KV-object staging. We describe the detailed API and execution protocol below.
\\

\noindent\textbf{\ding{203} Latency-hiding staging window.}
A larger internal-DRAM staging window can provide more slack to hide a slower SSD-backed path, but increasing device-side DRAM directly increases cost. Our prototype therefore keeps the internal-DRAM staging space bounded and instead scales the SSD-to-DRAM refill path. This design accommodates the growing prefix KV footprint across turns without proportionally scaling internal DRAM. Rather than using internal DRAM as a capacity cache for the accumulated prefix, \CXLHM streams KV objects through a fixed-depth staging window: upcoming objects are staged ahead of foreground remote reads, consumed by workers, and then released for later objects. As long as the refill pipeline keeps this window populated, bounded internal DRAM can support large and growing prefix KV caches. Thus, SSD bandwidth becomes a practical knob for sustaining prefetching, while internal DRAM remains a latency-hiding staging window rather than a cache sized to the total KV footprint.



\noindent\textbf{Read-Prioritized Write-Isolated KV Flushing.}
KV writes correspond to newly generated tokens and affect only future reuse, not the correctness of the current request. \CXLHM therefore isolates write handling from the latency-critical read path. The device reserves a small write-side buffer in internal DRAM and admits newly generated KV objects only when buffer space is available. If the buffer is full, the runtime can skip or retry the write later. Admitted KV objects are deferred and flushed asynchronously to the SSD-backed region in large batches, preferably when read pressure is low. This avoids interleaving background writes with foreground prefix reads on the SSD-backed path, preserving read-side prefetch bandwidth and reducing random read/write interference.

\noindent\textbf{User-Level Prefetch Coordination}
\CXLHM can receive object-level prefetch commands in the order that the objects will be consumed. For each prefetch command, \CXLHM stages the corresponding KV object from SSD-backed capacity into an internal-DRAM region. Later foreground remote reads can then be served from internal DRAM if the object has already been staged. To support KV-object prefetching in CXL-HM, we extend the user-directed prefetch interface of the \CXLHM{} prototype
with an explicit \texttt{issue--wait--release} protocol. The interface provides three object-level lifecycle APIs:

\begin{itemize}
\item \texttt{chm\_prefetch\_object(ptr, size)} prefetches an object into an internal-DRAM staging region and returns a request handle.

\item \texttt{chm\_prefetch\_wait(req)} waits until the prefetched object associated with the request handle is ready to be consumed.

\item \texttt{chm\_prefetch\_release(req)} releases the internal-DRAM staging region associated with the request after the object has been consumed.
\end{itemize}

For remote KV-cache reads, \CXLHM keeps up to a configured prefetch depth of KV objects in flight. Each prefetch stages an object from SSD-backed capacity into an internal-DRAM region. Once the object is consumed by a foreground remote read, the region is released back to the free pool. This bounded issue--consume--release protocol overlaps SSD-to-DRAM staging with remote reads while limiting internal DRAM usage.

\section{\sysname{}}

\subsection{\sysname Overview}

\begin{figure}[t]
\centering
\includegraphics[width=0.48\textwidth]{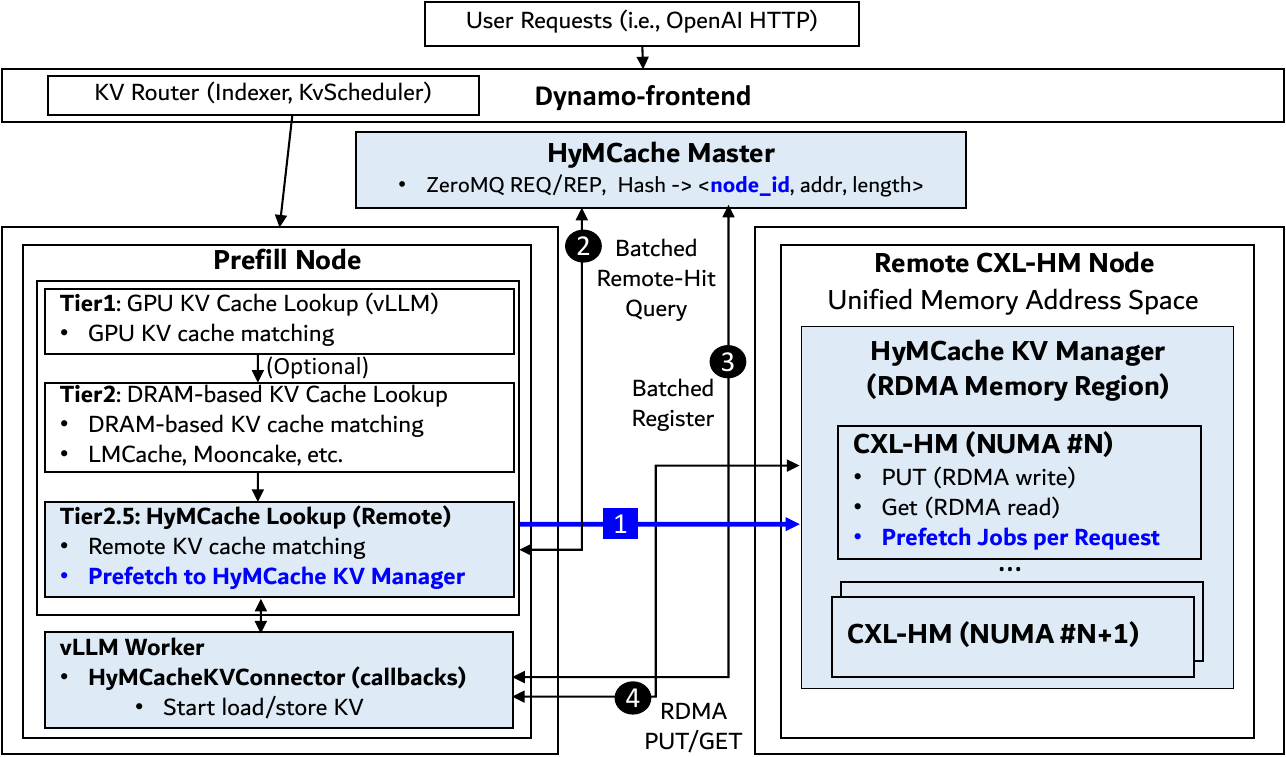}
\caption{\sysname overview and walkthrough example.
}
\label{fig:hycache-overview}
\end{figure}

To run within a practical LLM serving stack, \sysname integrates with Dynamo and vLLM, as shown in Figure~\ref{fig:hycache-overview}. On the hardware side, \sysname realizes the memory rack using
\CXLHM devices as the Tier-3 remote KV-cache tier. LLM workers can access this tier either through a CXL-based memory fabric, such as a CXL switch or CXL memory box, or through existing RDMA-capable datacenter networks. Our prototype focuses on the RDMA-based deployment because it can run on existing infrastructure without requiring a CXL fabric.

On the software side, the key distinguishing feature of \sysname is that it enables \CXLHM to provide remote-DRAM-like performance while exposing SSD-scale capacity. Specifically, \sysname aims to ensure that KV blocks residing in the SSD component of \CXLHM are prefetched in time to be present in \CXLHM’s internal DRAM when needed. To this end, we add four modules to the existing serving stack: Master, Lookup, KV Connector, and KV Manager. In the following, we describe the roles of the first three modules, while the KV Manager, which is part of \CXLHM, is described in Section~\ref{sec:CXLHM-KVManager}.

\noindent\textbf{\sysname Master.}
The \sysname Master module is responsible for metadata lookup and remote address resolution, which is done by maintaining global metadata for remote KV blocks across CXL-HM nodes. It maps each prefix/KV identifier to the remote access information needed for data transfer, including the target CXL-HM node, remote address, and length. 
The metadata footprint is small relative to the remote KV capacity and scales with the number of KV objects rather than raw capacity. With 16~MB KV blocks, 10~TB of remote KV capacity corresponds to approximately $6.5\times10^5$ KV objects. Assuming a compact 64~B metadata entry per object, covering the KV identifier, target CXL-HM node, offset, length, and state/version fields, the metadata footprint is about 40~MB for 10~TB of remote KV capacity.
 

More specifically, for writes, Master selects a target CXL-HM node at request granularity and returns the remote address where the KV block can be written. Keeping KV blocks from the same request on the same target node preserves ordered access opportunities, while different requests can be distributed across nodes for load balancing. After RDMA writes complete, the written KV blocks are registered with Master and added to the global metadata.

For reads, Master is queried with prefix/KV identifiers and Master returns the corresponding remote access information for blocks stored in \sysname. The returned metadata allows the worker-side KV Connector to issue RDMA reads directly to the target CXL-HM node. 

\noindent\textbf{\sysname Lookup.}
The Lookup module 
is responsible for early remote-hit matching, batched metadata lookup, and CXL-HM prefetch coordination before the vLLM worker reloads 
KV blocks into the GPU KV cache. After GPU and optional upper-tier cache lookup, Lookup 
immediately sends the currently available unmatched prefix blocks to \sysname{} Master, for up to 32 blocks per request by default, without waiting for the request to become full. Once the first remote hit identifies the target CXL-HM node, prefetch requests are issued for the remaining candidate suffix blocks on that node while Lookup continues to discover additional remote hits. The collected remote-hit metadata is then passed to the vLLM worker, so the latency-critical read path performs only data movement through batched RDMA reads.

\noindent\textbf{\sysname KV Connector.}
\sysname integrates with vLLM through a lightweight KV Connector attached to vLLM's existing KV-transfer callbacks. With \sysname disabled, vLLM would run unchanged. When enabled, KV Connector is invoked after vLLM's GPU prefix-cache lookup and any optional upper-tier DRAM-based KV-cache lookup, such as LMCache or Mooncake. For the remaining unmatched prefix blocks, KV Connector uses the remote-hit metadata produced by Lookup to issue RDMA reads into the GPU KV cache before attention, allowing the latency-critical read path to focus on RDMA data movement.
For newly generated KV blocks, KV Connector obtains write locations from Master, performs RDMA writes to the target CXL-HM node, and registers the written blocks in batch. To avoid blocking inference due to transient delays in remote KV retrieval, \sysname falls back to normal recomputation if the remote KV blocks are not ready within a 10~second timeout, following Mooncake's default configuration~\cite{qin2025mooncake}.

KV Connector also maintains a host-side CPU staging buffer for RDMA transfers, similar to existing KV-transfer systems such as LMCache~\cite{cheng2025lmcache}. This buffer is used as temporary transfer space between RDMA and the GPU KV cache, rather than as an additional cache tier. We use a 5~GB staging buffer by default, matching common practice in existing systems. For higher concurrency, this buffer should be scaled with the number of in-flight KV transfers.

\subsection{Remote KV Management on \CXLHM}
\label{sec:CXLHM-KVManager}
\noindent\textbf{\sysname KV Manager.}
The \sysname KV Manager module coordinates CXL-HM prefetching across concurrent requests and manages prefetch resources for each request. After receiving the remaining suffix block list, KV Manager performs remote-hit matching in the same request order as Lookup and invokes the CXL-HM prefetch API (Section~3.3) for matched KV objects. As both modules
process the same suffix list in the same order, the prefetch order is aligned with the subsequent RDMA read order.

To balance latency hiding and DRAM utilization, KV Manager dynamically allocates the prefetch window according to the prefix-hit footprint of each request while enforcing a bounded maximum prefetch window. Based on our prototype evaluation, allocating up to approximately 128~MB of prefetched KV data per request provides sustained performance. Consequently, KV Manager adjusts the prefetch window according to the KV-block size, up to eight blocks for 16~MB KV blocks and four blocks for 32~MB KV blocks. Requests with larger KV blocks or fewer prefix hits use a smaller prefetch window, down to double buffering when appropriate.

Each prefetched block is managed by KV Manager using the CXL-HM prefetch API and an opaque read token. For each matched KV object, KV Manager submits a prefetch request to the CXL-HM prefetch API, which stages the object into the CXL-HM's internal DRAM before the actual RDMA read. Once the prefetch request completes, KV Manager makes the corresponding descriptor available to the vLLM worker, indicating that the worker can fetch the object from the returned CXL-HM address. The worker then posts RDMA reads from the returned CXL-HM addresses into its local staging buffer and observes completion through its local RDMA completion queue. After the reads complete, the worker sends the consumed read tokens back to KV Manager, which releases the corresponding prefetched objects via the CXL-HM release API.



\noindent\textbf{Unified \CXLHM Address Space.}
On each remote \CXLHM{} node, each device appears as a CPU-less NUMA node, creating a one-to-one device-to-NUMA-node mapping. The \sysname{} KV Manager uses NUMA allocation interfaces~\cite{linux_numa_libnuma} to allocate memory from selected device-backed NUMA nodes for storing KV blocks, and registers the allocated memory as an RDMA memory region (MR). This RDMA registration provides an MR-level rkey, while per-block metadata records only the offset and length of each KV block within the registered memory region.

\sysname{} uses request-level NUMA assignment rather than page-level or fine-grained interleaving across devices. Across requests, the \sysname{} KV Manager selects NUMA nodes using simple policies such as round-robin or load-aware selection to balance capacity and bandwidth. After assigning a request to a NUMA node, the \sysname{} KV Manager allocates all prefix KV objects for that request from the selected node, placing them on the corresponding \CXLHM{} device. This enables ordered request-level prefetching through a single device-side staging path, without splitting the prefix stream across multiple \CXLHM{} devices. This NUMA-level placement remains internal to the \sysname{} KV Manager, while the rest of \sysname{} treats the remote \CXLHM{} capacity as a large memory space.



\noindent\textbf{Support for Heterogeneous KV Object Sizes.}
Because \sysname{} manages KV placement at object granularity, KV objects from different models, workers, and workloads can coexist in the same unified \CXLHM address space. Each request carries its model-specific KV object size, and its prefix list preserves the order in which KV objects are consumed. KV Manager uses this object size to determine the prefetch window and internal DRAM staging budget independently for each request. This allows concurrent LLM workloads with different KV object sizes to share the same \CXLHM tier without assuming a single global object size, while still balancing capacity and bandwidth across \CXLHM regions.

\subsection{A Walkthrough Example}

Figure~\ref{fig:hycache-overview} summarizes how \sysname{} coordinates \CXLHM prefetching, batched metadata operations, and RDMA data movement. On the read path, after GPU-side prefix-cache matching and optional upper-tier KV-cache lookup, the remaining prefix blocks are handled by the \sysname{} Lookup module, which \textcolor{blue}{\sqnum{1}} sends prefetch hints to the \sysname{} KV Manager so that upcoming KV objects can be staged in \CXLHM internal DRAM and \ding{203} 
issues a batched remote-hit query to the \sysname{} Master to obtain remote descriptors for matched KV objects. The matched descriptors are then passed to the vLLM worker, so the latency-critical read path performs only \ding{205} RDMA GETs into the GPU KV cache. On the write path, \ding{204} the worker sends a batched register request to the \sysname{} Master to obtain remote write descriptors, and \ding{205} writes the generated KV objects to the assigned \CXLHM addresses using RDMA PUTs.

\section{Experimental Setup}




\noindent \textbf{System configurations.}
We evaluate \sysname{} in a vLLM-based serving environment where workers access a remote CXL-HM KV tier over an RDMA-capable network. Our experiments cover two settings: PD-disaggregated serving, where prefill and decode execution are separated, and single-node serving, where they are co-located on the same worker.

For the PD-disaggregated setup, we use six Dell PowerEdge R770 servers, each equipped with dual Intel Xeon 6730 CPUs, 256~GB of DDR5 DRAM, and an NVIDIA A100 GPU with 80~GB of HBM. The remote \CXLHM tier is built using an FPGA-based \CXLHM{} prototype configured with 64~GB of device-side DRAM and two 1~TB Gen5$\times$4 backend SSDs.
exposed as a 2~TB shared memory tier to the host.
We map the serving roles as a 4P--1D--1S configuration: four servers run prefill workers, one server runs the decode worker and benchmark/control components, and one server runs the CXL-HM remote storage node. Each prefill server dedicates one 100~Gbps NIC port to the NIXL-based PD path and the other 100~Gbps NIC port to \sysname{} traffic. The decode and storage nodes each provide up to 200~Gbps of aggregate bandwidth to the prefill workers. This setup stresses KV-cache tier utilization under prefill-heavy, multi-worker access to reusable prefix KV blocks.

The single-node setup co-locates prefill and decode execution on one compute node and directly connects it to the CXL-HM remote KV tier. This removes PD scheduling effects and allows us to compare remote CXL-HM performance against local DRAM-based KV caching.

\begin{figure*}[t]
\centering
\includegraphics[width=0.9\textwidth]{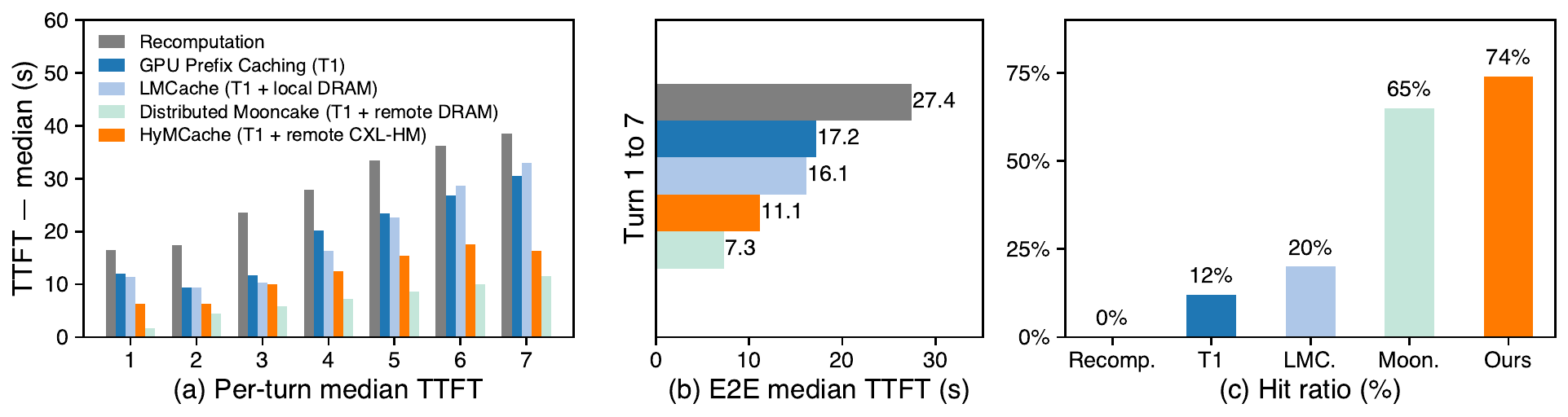}
\caption{TTFT in a 4P--1D--1S setup using Qwen2.5-32B. Recomputation performs no KV reuse, GPU prefix caching stores reusable KV blocks in GPU memory, LMCache stores them in local DRAM, and Mooncake stores them in remote distributed DRAM while requiring additional local DRAM as an RDMA transfer buffer.}
\label{fig:4node_qwen}
\end{figure*}

\sysname{} is implemented in vLLM 0.19.0 on top of the Dynamo v1.1.0 container image. Following Dynamo's LMCache integration model, we add a \sysname{} connector to vLLM's KV-transfer path during prefill execution. The connector coordinates remote KV lookup, prefetch requests, and RDMA-based KV transfer through the \sysname{} master and remote KV manager. When \sysname{} is disabled, vLLM follows its default execution path.

\noindent \textbf{Baselines.}
We compare \sysname{} against baselines that represent different levels of the KV-cache memory hierarchy:

\begin{itemize}
\item \textbf{Recomputation.}
No remote KV tier is used. Missing prefix KV blocks are recomputed through the default vLLM.

\item \textbf{GPU Prefix Caching (T1).}
vLLM reuses KV blocks that are already resident in the GPU KV cache.

\item \textbf{Local LMCache~\cite{cheng2025lmcache} (T1 + local DRAM).}
Each of the four workers is configured with 64~GB of local DRAM for KV caching.

\item \textbf{Distributed Mooncake~\cite{qin2025mooncake} (T1 + remote DRAM).}
Mooncake uses a distributed DRAM tier across prefill nodes, providing 1~TB of memory capacity for KV caching.

\item \textbf{Mooncake NVMe-oF~\cite{qin2025mooncake} (T1 + remote SSD).}
KV blocks are stored in a remote SSD-based KV tier and accessed through an RDMA-based NVMe-oF fabric. 

\item \textbf{\sysname{} (T1 + remote CXL-HM).}
\sysname{} serves reusable prefix KV blocks from a CXL-HM remote KV tier that combines internal DRAM with SSD-backed capacity.

\end{itemize}

Among the baselines, Distributed Mooncake and Mooncake NVMe-oF are the primary points of comparison for \sysname{}, as both expand their KV-cache capacity through a remote tier. However, the remote-tier components differ across these systems, leading to different performance and cost characteristics. We discuss the performance implications in the next section; here, we briefly estimate cost, focusing on the remote-tier components used in our experiments.

The FPGA-based \CXLHM prototype include an Agilex~7 development board~\cite{terasic_agilex7_iseries_price_2026}, 64~GB of device-side DRAM, and two Gen5 SSDs for a rough cost of \$10.5K--\$11.0K.
In comparison, Distributed Mooncake uses a 1~TB DDR5 RDIMM DRAM tier, which costs roughly \$30K--\$40K. Thus, even with FPGA prototyping overhead, \CXLHM is about 2.8--3.8$\times$ cheaper. The FPGA board alone accounts for roughly 91--95\% of the prototype cost, indicating that the estimate is dominated by prototyping overhead rather than by the SSDs or DRAM.
Mooncake NVMe-oF, on the other hand, can be considerably cheaper than \CXLHM by leveraging existing remote SSD infrastructure. However, practical deployments typically require integrating the remote NVMe devices through a host-side file system and storage software stack, introducing additional software overhead and latency, as discussed in Section~6.1. While these overheads may be reduced through DPU-based storage offloading, such deployments require additional accelerator hardware, which incurs additional costs.

\noindent\textbf{Models and Workloads.}
We evaluate \sysname{} using LLMs with different KV-cache footprints. Unless otherwise stated, we use FP16/BF16 KV cache and a vLLM block size of 128 tokens. We select Llama-3.1-8B and Qwen2.5-32B to cover different KV-object sizes used in our experiments. In our configuration, Llama-3.1-8B generates 16~MB KV blocks, while Qwen2.5-32B generates 32~MB KV blocks.

For workload generation, we use the synthetic benchmark provided by Dynamo AIPerf~\cite{aiperf-nividia,aiperf-git}, with mooncake trace~\cite{qin2025mooncake}, followed by the documented guideline from NVIDIA~\cite{aiperf-mooncake}.
AIPerf allows to benchmark LLM with controlled input and output length, prefix sharing ratio, request concurrency, and the number of prefill workers. We also use the LMSYS dataset~\cite{zheng2024lmsyschat1m} to evaluate \sysname{} under realistic multi-turn conversation patterns. To isolate the effect of KV-cache reuse, we set the maximum number of decode tokens to 1, minimizing decode-side computation. We report serving metrics including TTFT, end-to-end latency, remote-tier bandwidth, and KV-cache hit rate.

\section{Evaluation}



\subsection{PD-Disaggregated Serving Performance}
\subsubsection{Overall Performance Comparison}

Figure~\ref{fig:4node_qwen} shows the results under PD-disaggregated serving on Qwen2.5-32B using AIPerf synthetic inputs. This workload stresses cache capacity because the reusable prefix state quickly exceeds the capacity of GPU-only prefix caching and limited local DRAM caching, while generating approximately 1.5TB of reusable KV-cache footprint across turns.

Compared to Recomputation, GPU prefix caching, and LMCache-local caching, \sysname{} improves performance by retaining a larger fraction of reusable KV blocks and serving them from the remote CXL-HM tier.
Figure~\ref{fig:4node_qwen}(c) shows the cache-hit behavior at Turn~7. By this point, local cache baselines such as GPU prefix caching and LMCache local CPU caching exhibit low hit rates because their effective cache capacities are exceeded, and LRU eviction can break the contiguous prefix chain required for reuse.
Distributed Mooncake is more robust as it retains reusable prefix blocks in the remote DRAM tier, but its 1~TB capacity covers only about two-thirds of the 1.5~TB reusable KV footprint. In contrast, \sysname{} uses larger SSD-backed \CXLHM KV capacity to preserve more reusable prefix blocks, improving the hit rate by roughly 10\% over Mooncake and reducing prefill recomputation. This benefit is slightly limited by the read-prioritized \CXLHM{} policy, which delays write-buffer draining and skips more than 15\% of potential KV insertions under this workload.

Compared to \sysname{}, Distributed Mooncake with 1~TB of DRAM achieves roughly 30\% higher performance. This advantage comes primarily from its lower DRAM fetch latency and shared distributed-cache pool, which can reduce recomputation by preserving reusable KV blocks across workers in this workload. In contrast, \sysname{} incurs additional remote CXL-HM access latency and RDMA transfer overhead.

However, this gap should be interpreted in the context of the remote-tiers and their cost, where Distributed Mooncake is estimated to be much more costly, as discussed earlier.
More importantly, the performance gap narrows as the workload progresses to later turns. As the reusable KV footprint grows, capacity becomes increasingly important, allowing \sysname{} to retain reusable blocks that smaller caches would evict. This trend is visible in the figure where the gap between \sysname{} and the distributed-DRAM baseline decreases as the turn number increases.
Furthermore, the \CXLHM approach improves scalability because DRAM-only expansion is limited by DIMM slots, socket-level memory channels, and high capacity cost.
In contrast, \CXLHM can expand the remote KV tier by adding more \CXLHM devices or by increasing the SSD-backed capacity per device, with a much lower marginal cost per additional GB.

Overall, \sysname{} targets a practical middle point between high-cost remote DRAM and slow remote storage. It provides a much larger effective cache capacity than local memory, while avoiding the high recomputation cost of capacity-limited caches by approaching distributed-DRAM performance with a lower-cost \CXLHM tier.

\begin{figure}[t]
\centering
\includegraphics[width=0.99\columnwidth]{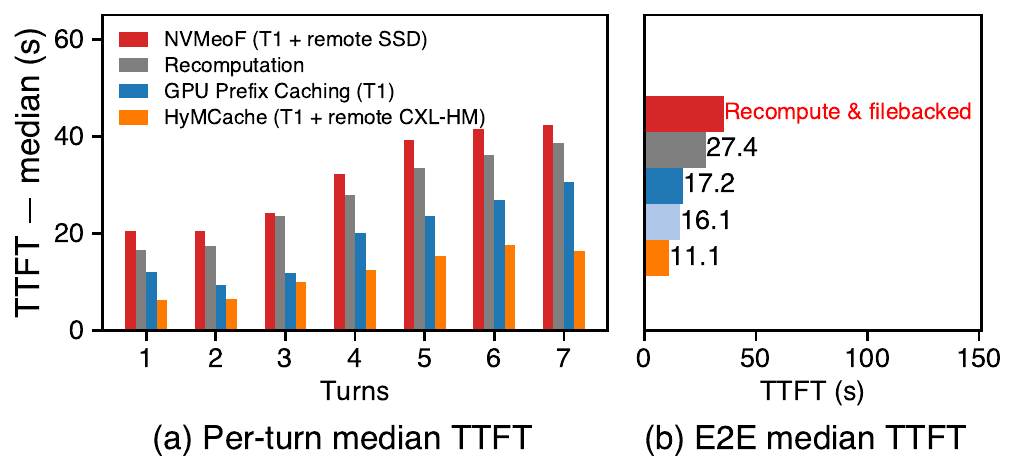}
\caption{Comparison with RDMA-based NVMe-oF Storage.}
\label{fig:4node_qwen_nvmeof}
\end{figure}

\begin{figure}[t]
\centering
\includegraphics[width=0.48\textwidth]{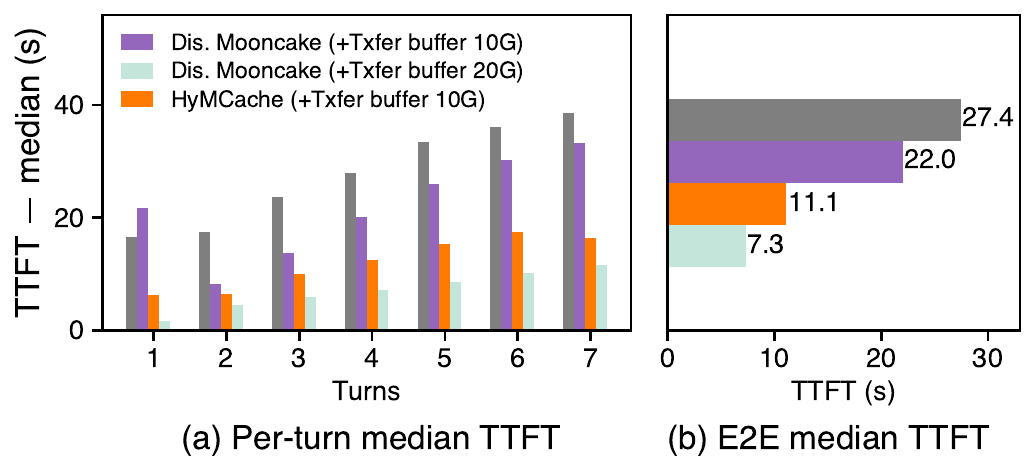}
\caption{Qwen2.5-32B serving performance with the Distributed Mooncake with 1~TB DRAM cache under different transfer-buffer sizes.}
\label{fig:4node_qwen_moon_transfer}
\end{figure}

\begin{figure*}[t]
\centering
\includegraphics[width=0.99\textwidth]{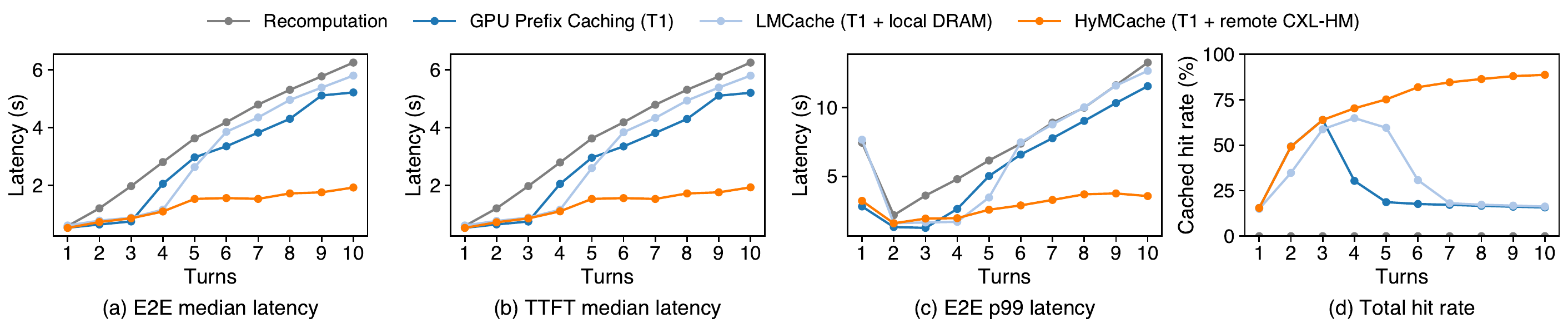}
\caption{End-to-end single-node serving performance comparing \sysname{} against baselines.}
\label{fig:singlenode_llama3}
\end{figure*}

\subsubsection{Comparison with NVMe-oF Remote Storage}
We now compare \sysname{} with an NVMe-oF-over-RDMA remote storage baseline using Mooncake. This baseline has the same setup as \CXLHM in that it uses the same ConnectX-6 RNIC and two Gen5$\times$4 SSDs and thus highlights the features of \CXLHM that have been tailored for multi-turn LLM. 

To support multiple prefill nodes in this baseline, the storage node first stripes the SSDs with RAID. To simplify our experimental setup, we then use LVM to partition the striped capacity and expose a separate NVMe-oF block volume to each prefill worker. Each remote volume is mounted with XFS, adding file-system overhead beyond the raw NVMe-oF transport.

Figure~\ref{fig:4node_qwen_nvmeof} shows that using an NVMe-oF storage tier substantially slows end-to-end serving compared with \sysname{}. In this configuration, remote KV fetches frequently exceed the timeout and fall back to normal recomputation, making the observed performance close to the recomputation baseline rather than benefiting from remote KV reuse. We attribute this degradation mainly to the software I/O path. We observe that the raw NVMe-oF block-device path reaches about 23~GB/s on each prefill node, close to the 200~Gbps network limit, whereas mounting XFS on top of the same NVMe-oF device reduces the measured sequential-read bandwidth to roughly 6.5~GB/s in our setup. 
This suggests that simply attaching high-bandwidth NVMe-oF storage is not sufficient for remote KV-cache serving; the storage path often requires additional optimization to deliver predictable bandwidth on the critical read path. Such optimizations may include careful I/O-path design, file-system bypass, DPU-assisted storage, or high-performance distributed file systems, but they add cost and system complexity. In contrast, \sysname{} leverages the memory-like semantics of CXL-HM to manage SSD-backed capacity as a TB-scale remote memory tier. By directly controlling KV-cache placement and prefetching over this tier, \sysname{} avoids placing a general-purpose file-system stack on the critical read path.




\begin{figure}[t]
\centering
\includegraphics[width=0.99\columnwidth]{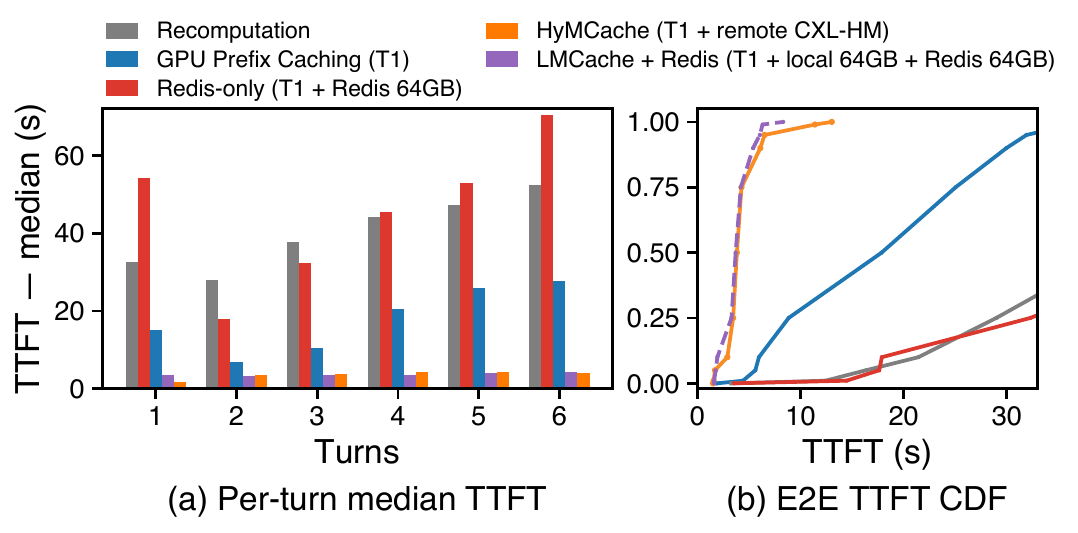}
\caption{Comparison with Redis-backed LMCache.
}

\label{fig:singleredis_qwen_cdf}
\vspace{-10pt}
\end{figure}

\subsubsection{Transfer-Buffer Sensitivity in Distributed DRAM Caching}
Although distributed DRAM caching is a high-performance reference point for remote KV reuse, its performance also depends on worker-side transfer-buffer capacity. For both Mooncake and \sysname{}, remote KV reuse follows a memory-based transfer path: KV blocks are received into CPU-side staging buffers and then copied into GPU-resident KV cache blocks. Sufficient staging space overlaps remote-to-host transfers with host-to-GPU copies, keeping remote reads ahead of GPU consumption. With limited staging space, remote reads are throttled by buffer availability, overlap is reduced, and cache-hit latency appears on the critical path. The required buffer also scales with KV-loading bandwidth; 100--400~Gbps links or multiple NICs need more staging space to keep KV reads in flight.

In the following set of experiments, we isolate this transfer-buffer effect and examine how much local host DRAM is needed to sustain memory-based remote KV reuse by varying the per-node transfer-buffer budget for both \sysname{} and Distributed Mooncake.
Figure~\ref{fig:4node_qwen_moon_transfer} shows their end-to-end serving performance. With the same 10~GB of additional host DRAM per node, Mooncake suffers from severe transfer-buffer pressure. Many remote KV insertions fail to complete in time, lowering effective KV reuse and often breaking the contiguous prefix chain in later turns. As a result, performance approaches the recomputation baseline, even with a longer transfer timeout.

\sysname{} is less sensitive to this constraint because it adopts an opportunistic remote-insertion policy. If the remote CXL-HM write buffer (Section~3.3) is full and cannot accept additional RDMA writes, \sysname{} skips the insertion and retries it in later turns rather than stalling the serving path. As the transfer-buffer budget increases, Mooncake recovers, showing that optimizing a distributed DRAM cache requires not only sufficient remote DRAM and RDMA bandwidth, but also additional local transfer-buffer DRAM at each worker. \sysname{}, however, shows similar performance beyond the 10~GB transfer-buffer budget.
\subsection{Single-Node Serving Performance}
\subsubsection{Overall Performance Comparison}

Figure~\ref{fig:singlenode_llama3} compares the single-node serving performance of \sysname{} against baseline caching methods. In this configuration, prefill and decode execution are co-located on the same LLM worker, allowing us to isolate the latency and hit-rate effects of remote KV caching. We evaluate Llama-3.1-8B using the LMSYS dataset with 512 multi-turn conversations, a concurrency of 32, and up to 10 turns per conversation. We set the maximum number of decode tokens to 1 to focus on prefix reuse and prefill-side latency.

The key advantage of \sysname{} is that it extends KV-cache capacity with a remote \CXLHM{} tier, enabling large-capacity prefix reuse and reducing recomputation. Recomputation shows steadily increasing latency as turns progress because the accumulated prefix becomes longer at each turn. GPU prefix caching and LMCache local CPU caching improve performance in early turns, but their benefits decrease once the accumulated prefix working set exceeds their effective cache capacity. After this point, GPU prefix caching begins to lose effective reuse around Turn~3, and LMCache local CPU caching around Turn~4. 
This is because LRU eviction can remove earlier prefix blocks as the conversation grows. Since prefix reuse depends on a contiguous chain of cached blocks from the beginning of the prompt, evicting an early block truncates the reusable prefix, and later cached blocks no longer translate into effective prefix hits. Consequently, the cache hit rate drops sharply, more of the accumulated prefix is recomputed, and latency gradually approaches the recomputation baseline. In contrast, \sysname{} preserves reusable prefix KV blocks in the \CXLHM tier and maintains a hit rate close to 90\%, avoiding this recomputation latency growth across turns.


\subsubsection{Comparison with LMCache Remote Redis Tier.}
We further compare \sysname{} against LMCache configured with a remote Redis tier~\cite{lmcache_redis_docs}. 
We evaluate two Redis configurations: Redis-only 64~GB remote caching and LMCache with a 64~GB local DRAM cache plus an additional 64~GB Redis-backed remote cache tier accessed over TCP. For this experiment, we use Qwen2.5-32B and choose a workload footprint that fits within the combined 128~GB cache capacity of the LMCache and Redis configuration.

Figure~\ref{fig:singleredis_qwen_cdf}(a) shows serving performance across Turns~1--6. 
When Redis is used alone as the remote cache tier, frequent store failures break the reusable prefix chain and cause requests to fall back to recomputation. Adding LMCache local DRAM caching makes the configuration more stable, as prefix KV blocks can be retained and reused through the combined local and Redis-backed cache.

Figure~\ref{fig:singleredis_qwen_cdf}(b) shows the latency CDF at Turn~6.
Although LMCache and Redis benefits from a memory-backed local/remote cache combination, \sysname{} maintains a similar latency distribution by preserving reusable prefix KV blocks in the remote CXL-HM-backed tier. This shows that \sysname{} can sustain effective prefix reuse using remote CXL-HM capacity, even when compared with a memory-based cache hierarchy.

\begin{figure}[t]
\centering
\includegraphics[width=0.48\textwidth]{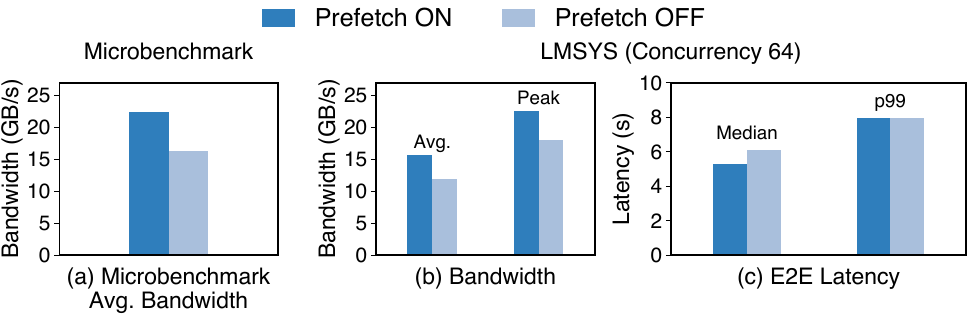}
\caption{Performance impact of \sysname{} prefetching}
\label{fig:prefetch-on-off}
\vspace{-10pt}
\end{figure}

\subsubsection{Impact of \sysname{} Prefetching}
We analyze the impact of CXL-HM prefetching by comparing execution with and without \sysname{} prefetching. To better isolate the benefit of prefetching under read-intensive access patterns, we first design a microbenchmark that issues heavy random-read requests to the remote \CXLHM{} tier managed by \sysname{}. As shown in Figure~\ref{fig:prefetch-on-off}(a), enabling prefetching improves performance by 37\% compared to the prefetch-off configuration.

Figures~\ref{fig:prefetch-on-off}(b) and~\ref{fig:prefetch-on-off}(c) further evaluate prefetching using the LMSYS dataset. To create a read-heavy scenario, we select requests with input lengths longer than 2K characters and run 512 conversations with a concurrency of 64. We report the results at Turn~6, where the external cache hit rate exceeds 90\%. Figure~\ref{fig:prefetch-on-off}(b) shows the average and peak remote KV loading bandwidth. With prefetching, the peak bandwidth approaches the 200~Gbps network limit, while the average bandwidth improves by more than 35\% over the prefetch-off case. As a result, Figure~\ref{fig:prefetch-on-off}(c) shows that prefetching reduces end-to-end execution time by 14\%. These results show that \sysname{} prefetching is effective for read-heavy multi-turn workloads because it pipelines data staging within the \CXLHM{} tier and RDMA-based KV transfer to the worker.

\section{Related work}

\noindent \textbf{CXL-based Memory Expansion and Pooling.}
CXL has been explored as a substrate for memory expansion, pooling, and disaggregation. Prior work has studied CXL-based memory expansion for improving memory capacity beyond local DIMM channels~\cite{gouk2023memory,Jang2023atc,aguilera2023memory,zhou2024lightwsp,giannoula2023daemon}, as well as CXL-aware tiered memory management for heterogeneous memory systems~\cite{almaruf2023tpp}. Due to the limited availability of commercial CXL hardware, many early studies relied on simulation, emulation, or software-based modeling to evaluate CXL memory behavior~\cite{esmaili_dokht2024mess,fridman2023cxl,arif2022exploiting}. More recent commercial and prototype systems demonstrate several CXL design points, including DRAM-based CXL memory expanders~\cite{skhynix_cmm_ddr5,samsung_cmm_d}, switch-based CXL memory pools and appliances~\cite{samsung_cmm_b,skhynix_niagara2,beluga_cxl_kvcache}, and \CXLHM devices that combine internal DRAM with SSD-backed capacity~\cite{samsung_cmmh_hotstorage,jang2026itmeinferencetieredmemory}.

\sysname{} differs from these efforts in both target workload and system interface. Rather than using CXL memory as a generic memory expansion tier,
\sysname builds a CXL memory rack for remote KV caching in
multi-turn LLM serving. In this setting, KV accesses are large, read-heavy, and predictable because prefix KV blocks are repeatedly reused across turns. \sysname{} exploits this property by coordinating serving-level KV metadata with \CXLHM prefetching and device-side DRAM staging. This allows SSD-backed \CXLHM capacity to act as a cost-efficient remote KV tier while keeping latency-critical reused KV blocks on the fast internal DRAM path.

\noindent \textbf{Multi-tier KV Cache and Prefetching.}
The large memory footprint of KV caches has motivated work on memory tiering and cache reuse for LLM inference. General inference systems such as FlexGen~\cite{sheng2023flexgen} and DeepSpeed Inference~\cite{aminabadi2022deepspeed} improve throughput by partitioning model data and intermediate states across GPU, CPU, and NVMe storage, while LLM in a Flash~\cite{alizadeh2023llmflash} and PowerInfer~\cite{song2023powerinfer} further explore flash-based offloading for memory-constrained inference. For KV-cache management, PagedAttention~\cite{kwon2023vllm}, Pensieve~\cite{yu2023pensieve}, and CachedAttention~\cite{gao2024cachedattention} improve paging, placement, and context reuse, while LMCache~\cite{cheng2025lmcache} and Mooncake~\cite{qin2025mooncake} reduce recomputation through KV sharing and disaggregated KV-cache architectures.

\sysname{} is complementary to these software-level systems but targets a different layer of the hierarchy. Rather than proposing another GPU/CPU/NVMe cache policy, \sysname{} provides a \CXLHM based remote memory backend for large reusable KV states. Its key distinction is that it couples serving-level knowledge of prefix KV reuse with device-level DRAM staging: the serving system identifies reusable KV blocks before they are consumed and conveys this information to \CXLHM through prefetch commands. This allows \sysname{} to use SSD-backed capacity for scale while staging latency-critical KV reads in device-side DRAM, making \CXLHM an efficient lower-tier backend underneath existing KV-cache management policies.

\noindent \textbf{DPU-based Storage and Remote KV Caching.}
DPU and SmartNIC based storage disaggregation has enabled high-density remote flash systems and JBOF-style storage appliances~\cite{supermicro_jbof}. Prior work and products such as LEED~\cite{guo2023leed}, Gimbal~\cite{gimbal_sigcomm21}, NVMe-oF target offload~\cite{nvmeof_target_offload_perf,nvme_offload_whitepaper}, Ditto~\cite{ditto_sosp23}, and FORD~\cite{ford_fast22} reduce host CPU overhead or improve remote storage access by offloading indexing, storage processing, and RDMA/NVMe-oF data movement.

\sysname{} takes a different approach. Rather than optimizing a storage-oriented path to remote flash, \sysname{} builds a memory-addressable remote KV tier using \CXLHM In this design, SSD-backed capacity is exposed as remote memory, while device-side DRAM is managed as a staging layer for latency-critical KV reads. This avoids relying on DPU cores or storage-node software to mediate every SSD-resident KV access, and instead uses \CXLHM as a cost-efficient remote KV-cache backend for predictable KV reuse in multi-turn LLM serving.

\section{Conclusion}
This paper presents \sysname, a CXL memory rack for multi-turn LLM
serving. We build the rack using cost-efficient CXL-hybrid memory,
which exposes large SSD-backed capacity through a memory interface
while using device-side DRAM for latency-critical staging. Leveraging the read-dominant, predictable, and append-only access patterns of multi-turn KV caches, \sysname{} rethinks DRAM management within \CXLHM{} through request-level prefix prefetching and opportunistic write buffering. This allows latency-critical reusable KV blocks to be staged in device DRAM while relying on SSD-backed capacity for TB-scale KV reuse. Our evaluation on a real \CXLHM{} prototype under both single-aggregator and PD-disaggregated serving shows that \sysname{} enables scalable and cost-efficient remote KV-cache capacity for future LLM serving systems.


\bibliographystyle{ACM-Reference-Format}
\bibliography{ref}

\end{document}